\documentclass{fac}

\usepackage{graphicx}
\usepackage{amssymb}
\usepackage{amsfonts}
\usepackage{amsmath}
\usepackage{dsfont}
\usepackage{MnSymbol}
\usepackage{mathtools}
\usepackage{epsfig}
\usepackage{epstopdf}
\usepackage{tikz}
\usepackage{amsthm}
\ifprodtf \else \usepackage{latexsym}\fi

\newcommand\black{\ensuremath{\blacktriangleright}}
\newcommand\white{\ensuremath{\vartriangleright}}

\newif\ifamsfontsloaded
\ifprodtf
  \newcommand\whbl{\white\kern-.1em--\kern-.1em\black}
  \newcommand\blwh{\black\kern-.1em--\kern-.1em\white}
  \newcommand\blbl{\black\kern-.1em--\kern-.1em\black}
  \newcommand\whwh{\white\kern-.1em--\kern-.1em\white}
  \amsfontsloadedtrue
\else
  \checkfont{msam10}
  \iffontfound
    \IfFileExists{amssymb.sty}
      {\usepackage{amssymb}\amsfontsloadedtrue
       \newcommand\whbl{\white\kern-.125em--\kern-.125em\black}%
       \newcommand\blwh{\black\kern-.125em--\kern-.125em\white}%
       \newcommand\blbl{\black\kern-.125em--\kern-.125em\black}%
       \newcommand\whwh{\white\kern-.125em--\kern-.125em\white}}
      {}
  \fi
\fi

\newtheorem{theorem}{Theorem}[section]
\newtheorem{definition}[theorem]{Definition}

\title[Draft of Truly Concurrent Bisimilarities are Game Equivalent]
      {Truly Concurrent Bisimilarities are Game Equivalent}

\author[Yong Wang]
    {Yong Wang\\
     College of Computer Science and Technology,\\
     Faculty of Information Technology,\\
     Beijing University of Technology, Beijing, China\\
     }

\correspond{Yong Wang, Pingleyuan 100, Chaoyang District, Beijing, China.
            e-mail: wangy@bjut.edu.cn}

\pubyear{2019}
\pagerange{\pageref{firstpage}--\pageref{lastpage}}

\begin{document}
\label{firstpage}

\makecorrespond

\maketitle

\begin{abstract}
We design games for truly concurrent bisimilarities, including strongly truly concurrent bisimilarities and branching truly concurrent bisimilarities, such as pomset bisimilarities, step bisimilarities, history-preserving bisimilarities and hereditary history-preserving bisimilarities.
\end{abstract}


\section{Introduction}{\label{int}}

Game theory has been widely used to interpret the nature of the world. The combination of game theory and (computational) logic \cite{LIG} always exists two ways.

One is to give game theory a logic basis, such as game logic \cite{GL1} \cite{GL2} \cite{GL3}, game algebras \cite{BAG1} \cite{BAG2}, algebras \cite{CG4} for concurrent games
\cite{CG1} \cite{CG2} \cite{CG3}.

the other is to use game theory to interpret computational logic, such as the well-known game semantics \cite{PCF} \cite{PCF2} \cite{PCF3} \cite{MIL} \cite{Algol}, in which game theory acts as a foundational semantics bases to understand the behaviors of computer programming language.

Recently, there are some work on interpreting bisimilarities by use of game theory \cite{GBA1} \cite{GBA2}. Following these work, we give truly concurrent bisimilarities a game theory interpretation. This work is organized as follows. In section \ref{gpb}, \ref{gsb}, \ref{ghb}, \ref{ghhb}, we design games for pomset bisimilarites, step bisimilarities, history-preserving bisimilarities, and hereditary history-preserving bisimilarities, respectively. And finally, in section \ref{con}, we conclude this paper.

\section{Games for Pomset Bisimulations}\label{gpb}

\begin{definition}[Prime event structure with silent event(\cite{APTC})]\label{PES}
Let $\Lambda$ be a fixed set of labels, ranged over $a,b,c,\cdots$ and $\tau$. A ($\Lambda$-labelled) prime event structure with silent event $\tau$ is a tuple $\mathcal{E}=\langle \mathbb{E}, \leq, \sharp, \lambda\rangle$, where $\mathbb{E}$ is a denumerable set of events, including the silent event $\tau$. Let $\hat{\mathbb{E}}=\mathbb{E}\backslash\{\tau\}$, exactly excluding $\tau$, it is obvious that $\hat{\tau^*}=\epsilon$, where $\epsilon$ is the empty event. Let $\lambda:\mathbb{E}\rightarrow\Lambda$ be a labelling function and let $\lambda(\tau)=\tau$. And $\leq$, $\sharp$ are binary relations on $\mathbb{E}$, called causality and conflict respectively, such that:

\begin{enumerate}
  \item $\leq$ is a partial order and $\lceil e \rceil = \{e'\in \mathbb{E}|e'\leq e\}$ is finite for all $e\in \mathbb{E}$. It is easy to see that $e\leq\tau^*\leq e'=e\leq\tau\leq\cdots\leq\tau\leq e'$, then $e\leq e'$.
  \item $\sharp$ is irreflexive, symmetric and hereditary with respect to $\leq$, that is, for all $e,e',e''\in \mathbb{E}$, if $e\sharp e'\leq e''$, then $e\sharp e''$.
\end{enumerate}

Then, the concepts of consistency and concurrency can be drawn from the above definition:

\begin{enumerate}
  \item $e,e'\in \mathbb{E}$ are consistent, denoted as $e\frown e'$, if $\neg(e\sharp e')$. A subset $X\subseteq \mathbb{E}$ is called consistent, if $e\frown e'$ for all $e,e'\in X$.
  \item $e,e'\in \mathbb{E}$ are concurrent, denoted as $e\parallel e'$, if $\neg(e\leq e')$, $\neg(e'\leq e)$, and $\neg(e\sharp e')$.
\end{enumerate}
\end{definition}

\begin{definition}[Configuration(\cite{APTC})]
Let $\mathcal{E}$ be a PES. A (finite) configuration in $\mathcal{E}$ is a (finite) consistent subset of events $C\subseteq \mathcal{E}$, closed with respect to causality (i.e. $\lceil C\rceil=C$). The set of finite configurations of $\mathcal{E}$ is denoted by $\mathcal{C}(\mathcal{E})$. We let $\hat{C}=C\backslash\{\tau\}$.
\end{definition}

A consistent subset of $X\subseteq \mathbb{E}$ of events can be seen as a pomset. Given $X, Y\subseteq \mathbb{E}$, $\hat{X}\sim \hat{Y}$ if $\hat{X}$ and $\hat{Y}$ are isomorphic as pomsets. In the following of the paper, we say $C_1\sim C_2$, we mean $\hat{C_1}\sim\hat{C_2}$.

\subsection{Games for Strong Pomset Bisimulation}

\begin{definition}[Pomset transitions(\cite{APTC})]
Let $\mathcal{E}$ be a PES and let $C\in\mathcal{C}(\mathcal{E})$, and $\emptyset\neq X\subseteq \mathbb{E}$, if $C\cap X=\emptyset$ and $C'=C\cup X\in\mathcal{C}(\mathcal{E})$, then $C\xrightarrow{X} C'$ is called a pomset transition from $C$ to $C'$.
\end{definition}

\begin{definition}[Pomset bisimulation(\cite{APTC})]\label{PSB}
Let $\mathcal{E}_1$, $\mathcal{E}_2$ be PESs. A pomset bisimulation is a relation $R\subseteq\mathcal{C}(\mathcal{E}_1)\times\mathcal{C}(\mathcal{E}_2)$, such that if $(C_1,C_2)\in R$, and $C_1\xrightarrow{X_1}C_1'$ then $C_2\xrightarrow{X_2}C_2'$, with $X_1\subseteq \mathbb{E}_1$, $X_2\subseteq \mathbb{E}_2$, $X_1\sim X_2$ and $(C_1',C_2')\in R$, and vice-versa. We say that $\mathcal{E}_1$, $\mathcal{E}_2$ are pomset bisimilar, written $\mathcal{E}_1\sim_p\mathcal{E}_2$, if there exists a pomset bisimulation $R$, such that $(\emptyset,\emptyset)\in R$.
\end{definition}

The Ehrenfeucht-Fra\"{i}ss\'{e} game for strong pomset bisimulation, in which two players called Spoiler and Duplicator exist, is as follows.

\begin{definition}[Game for pomset bisimulation]
A (strong) pomset bisimulation game on $\mathbb{E}$ is played on an arena of Spoiler-owned configurations $[(C_1,C_2)]$ and Duplicator-owned configurations $\langle(C_1,C_2),c\rangle$, where $(C_1,C_2)\in S\times S$ the set of positions, and $c\in A\times S$ the set of pending challenges, with:
\begin{itemize}
  \item Spoiler moves from a configuration $[(C_1,C_2)]$ by:
  \begin{enumerate}
    \item selecting $C_1\xrightarrow{X_1}C_1'$ and moving to $\langle(C_1,C_2),(X_1,C_1')\rangle$ with $X_1\subseteq \mathbb{E}_1$, or
    \item selecting $C_2\xrightarrow{X_2}C_2'$ and moving to $\langle (C_2,C_1),(X_2,C_2')\rangle$ with $X_2\subseteq\mathbb{E}_2$;
    \item $X_1\sim X_2$;
  \end{enumerate}
  \item Duplicator responds from a configuration $\langle (C_3,C_4), (X_1,C_3')$ by playing $C_4\xrightarrow{X_2}C_4'$ and continuing in configuration $[(C_3',C_4')]$.
\end{itemize}

If games starting in a configuration $[(C_1,C_2)]$ is won by Duplicator, we write $C_1\equiv_p C_2$.
\end{definition}

\begin{theorem}
$C_1\sim_p C_2$ iff $C_1\equiv_p C_2$.
\end{theorem}

\subsection{Games for Branching Pomset Bisimulation}

\begin{definition}[Branching pomset bisimulation(\cite{APTC})]\label{BPB}
Assume a special termination predicate $\downarrow$, and let $\surd$ represent a state with $\surd\downarrow$. Let $\mathcal{E}_1$, $\mathcal{E}_2$ be PESs. A branching pomset bisimulation is a relation $R\subseteq\mathcal{C}(\mathcal{E}_1)\times\mathcal{C}(\mathcal{E}_2)$, such that:
 \begin{enumerate}
   \item if $(C_1,C_2)\in R$, and $C_1\xrightarrow{X}C_1'$ then
   \begin{itemize}
     \item either $X\equiv \tau^*$, and $(C_1',C_2)\in R$;
     \item or there is a sequence of (zero or more) $\tau$-transitions $C_2\xrightarrow{\tau^*} C_2^0$, such that $(C_1,C_2^0)\in R$ and $C_2^0\xRightarrow{X}C_2'$ with $(C_1',C_2')\in R$;
   \end{itemize}
   \item if $(C_1,C_2)\in R$, and $C_2\xrightarrow{X}C_2'$ then
   \begin{itemize}
     \item either $X\equiv \tau^*$, and $(C_1,C_2')\in R$;
     \item or there is a sequence of (zero or more) $\tau$-transitions $C_1\xrightarrow{\tau^*} C_1^0$, such that $(C_1^0,C_2)\in R$ and $C_1^0\xRightarrow{X}C_1'$ with $(C_1',C_2')\in R$;
   \end{itemize}
   \item if $(C_1,C_2)\in R$ and $C_1\downarrow$, then there is a sequence of (zero or more) $\tau$-transitions $C_2\xrightarrow{\tau^*}C_2^0$ such that $(C_1,C_2^0)\in R$ and $C_2^0\downarrow$;
   \item if $(C_1,C_2)\in R$ and $C_2\downarrow$, then there is a sequence of (zero or more) $\tau$-transitions $C_1\xrightarrow{\tau^*}C_1^0$ such that $(C_1^0,C_2)\in R$ and $C_1^0\downarrow$.
 \end{enumerate}

We say that $\mathcal{E}_1$, $\mathcal{E}_2$ are branching pomset bisimilar, written $\mathcal{E}_1\approx_{bp}\mathcal{E}_2$, if there exists a branching pomset bisimulation $R$, such that $(\emptyset,\emptyset)\in R$.
\end{definition}

The Ehrenfeucht-Fra\"{i}ss\'{e} game for branching pomset bisimulation, in which two players called Spoiler and Duplicator exist, is as follows. Note that the game is limited to no occurrence of infinite $\tau$-loops.

\begin{definition}[Game for branching pomset bisimulation]
A limited branching pomset bisimulation game on $\mathbb{E}$ is played on an arena of Spoiler-owned configurations $[(C_1,C_2)]$ and Duplicator-owned configurations $\langle(C_1,C_2),c\rangle$, where $(C_1,C_2)\in S\times S$ the set of positions, and $c\in A\times S$ the set of pending challenges, with:
\begin{itemize}
  \item Spoiler moves from a configuration $[(C_1,C_2)]$ by:
  \begin{enumerate}
    \item selecting $C_1\xrightarrow{X}C_1'$ and moving to $\langle(C_1,C_2),(X,C_1')\rangle$ with $X\subseteq \mathbb{E}$, or
    \item selecting $C_2\xrightarrow{X}C_2'$ and moving to $\langle (C_2,C_1),(X,C_2')\rangle$ with $X\subseteq\mathbb{E}$;
  \end{enumerate}
  \item Duplicator responds from a configuration $\langle (C_3,C_4), (X,C_3')$ by:
  \begin{enumerate}
    \item If $X=\tau*$, continuing in the configuration $[(C_3',C_4)]$, or
    \item playing $C_4\xrightarrow{X}C_4'$ and continuing in configuration $[(C_3',C_4')]$ if available, or
    \item playing $C_4\xrightarrow{\tau*}C_4'$ and continuing in configuration $[(C_3,C_4')]$ if possible.
  \end{enumerate}
\end{itemize}

If games starting in a configuration $[(C_1,C_2)]$ is won by Duplicator, we write $C_1\equiv_{bp} C_2$.
\end{definition}

\begin{theorem}
$C_1\approx_{bp} C_2$ iff $C_1\equiv_{bp} C_2$.
\end{theorem}

\section{Games for Step Bisimulations}\label{gsb}

\subsection{Games for Strong Step Bisimulation}

\begin{definition}[Step transitions(\cite{APTC})]
Let $\mathcal{E}$ be a PES and let $C\in\mathcal{C}(\mathcal{E})$, and $\emptyset\neq X\subseteq \mathbb{E}$, if $C\cap X=\emptyset$ and $C'=C\cup X\in\mathcal{C}(\mathcal{E})$, then $C\xrightarrow{X} C'$ is called a pomset transition from $C$ to $C'$. When the events in $X$ are pairwise concurrent, we say that $C\xrightarrow{X}C'$ is a step.
\end{definition}

\begin{definition}[Step bisimulation(\cite{APTC})]\label{SB}
By replacing pomset transitions with steps, we can get the definition of step bisimulation. When PESs $\mathcal{E}_1$ and $\mathcal{E}_2$ are step bisimilar, we write $\mathcal{E}_1\sim_s\mathcal{E}_2$.
\end{definition}

The Ehrenfeucht-Fra\"{i}ss\'{e} game for strong step bisimulation, in which two players called Spoiler and Duplicator exist, is as follows.

\begin{definition}[Game for step bisimulation]
A (strong) step bisimulation game on $\mathbb{E}$ is played on an arena of Spoiler-owned configurations $[(C_1,C_2)]$ and Duplicator-owned configurations $\langle(C_1,C_2),c\rangle$, where $(C_1,C_2)\in S\times S$ the set of positions, and $c\in A\times S$ the set of pending challenges, with:
\begin{itemize}
  \item Spoiler moves from a configuration $[(C_1,C_2)]$ by:
  \begin{enumerate}
    \item selecting $C_1\xrightarrow{X_1}C_1'$ and moving to $\langle(C_1,C_2),(X_1,C_1')\rangle$ with $X_1\subseteq \mathbb{E}_1$, and all $e\in X_1$ are pairwise concurrent, or
    \item selecting $C_2\xrightarrow{X_2}C_2'$ and moving to $\langle (C_2,C_1),(X_2,C_2')\rangle$ with $X_2\subseteq\mathbb{E}_2$, and all $e\in X_2$ are pairwise concurrent;
    \item $X_1\sim X_2$;
  \end{enumerate}
  \item Duplicator responds from a configuration $\langle (C_3,C_4), (X_1,C_3')$ by playing $C_4\xrightarrow{X_2}C_4'$ and continuing in configuration $[(C_3',C_4')]$.
\end{itemize}

If games starting in a configuration $[(C_1,C_2)]$ is won by Duplicator, we write $C_1\equiv_s C_2$.
\end{definition}

\begin{theorem}
$C_1\sim_s C_2$ iff $C_1\equiv_s C_2$.
\end{theorem}

\subsection{Games for Branching Step Bisimulation}

\begin{definition}[Branching step bisimulation(\cite{APTC})]\label{BSB}
For $\mathcal{E}_1\approx_{bp}\mathcal{E}_2$, by replacing pomset transitions with steps, we can get the definition of branching step bisimulation. When PESs $\mathcal{E}_1$ and $\mathcal{E}_2$ are branching step bisimilar, we write $\mathcal{E}_1\approx_{bs}\mathcal{E}_2$.
\end{definition}

The Ehrenfeucht-Fra\"{i}ss\'{e} game for branching step bisimulation, in which two players called Spoiler and Duplicator exist, is as follows. Note that the game is limited to no occurrence of infinite $\tau$-loops.

\begin{definition}[Game for branching step bisimulation]
A limited branching step bisimulation game on $\mathbb{E}$ is played on an arena of Spoiler-owned configurations $[(C_1,C_2)]$ and Duplicator-owned configurations $\langle(C_1,C_2),c\rangle$, where $(C_1,C_2)\in S\times S$ the set of positions, and $c\in A\times S$ the set of pending challenges, with:
\begin{itemize}
  \item Spoiler moves from a configuration $[(C_1,C_2)]$ by:
  \begin{enumerate}
    \item selecting $C_1\xrightarrow{X}C_1'$ and moving to $\langle(C_1,C_2),(X,C_1')\rangle$ with $X\subseteq \mathbb{E}$, and all $e\in X$ are pairwise concurrent, or
    \item selecting $C_2\xrightarrow{X}C_2'$ and moving to $\langle (C_2,C_1),(X,C_2')\rangle$ with $X\subseteq\mathbb{E}$, and all $e\in X$ are pairwise concurrent;
  \end{enumerate}
  \item Duplicator responds from a configuration $\langle (C_3,C_4), (X,C_3')$ by:
  \begin{enumerate}
    \item If $X=\tau*$, continuing in the configuration $[(C_3',C_4)]$, or
    \item playing $C_4\xrightarrow{X}C_4'$ and continuing in configuration $[(C_3',C_4')]$ if available, or
    \item playing $C_4\xrightarrow{\tau*}C_4'$ and continuing in configuration $[(C_3,C_4')]$ if possible.
  \end{enumerate}
\end{itemize}

If games starting in a configuration $[(C_1,C_2)]$ is won by Duplicator, we write $C_1\equiv_{bs} C_2$.
\end{definition}

\begin{theorem}
$C_1\approx_{bs} C_2$ iff $C_1\equiv_{bs} C_2$.
\end{theorem}

\section{Games for History-preserving Bisimulations}\label{ghb}

\subsection{Games for Strong History-preserving Bisimulation}

\begin{definition}[Posetal product(\cite{APTC})]
Given two PESs $\mathcal{E}_1$, $\mathcal{E}_2$, the posetal product of their configurations, denoted $\mathcal{C}(\mathcal{E}_1)\overline{\times}\mathcal{C}(\mathcal{E}_2)$, is defined as

$$\{(C_1,f,C_2)|C_1\in\mathcal{C}(\mathcal{E}_1),C_2\in\mathcal{C}(\mathcal{E}_2),f:C_1\rightarrow C_2 \textrm{ isomorphism}\}.$$

A subset $R\subseteq\mathcal{C}(\mathcal{E}_1)\overline{\times}\mathcal{C}(\mathcal{E}_2)$ is called a posetal relation. We say that $R$ is downward closed when for any $(C_1,f,C_2),(C_1',f',C_2')\in \mathcal{C}(\mathcal{E}_1)\overline{\times}\mathcal{C}(\mathcal{E}_2)$, if $(C_1,f,C_2)\subseteq (C_1',f',C_2')$ pointwise and $(C_1',f',C_2')\in R$, then $(C_1,f,C_2)\in R$.

For $f:X_1\rightarrow X_2$, we define $f[x_1\mapsto x_2]:X_1\cup\{x_1\}\rightarrow X_2\cup\{x_2\}$, $z\in X_1\cup\{x_1\}$,(1)$f[x_1\mapsto x_2](z)=
x_2$,if $z=x_1$;(2)$f[x_1\mapsto x_2](z)=f(z)$, otherwise. Where $X_1\subseteq \mathbb{E}_1$, $X_2\subseteq \mathbb{E}_2$, $x_1\in \mathbb{E}_1$, $x_2\in \mathbb{E}_2$.
\end{definition}

\begin{definition}[History-preserving bisimulation(\cite{APTC})]\label{HPB}
A history-preserving (hp-) bisimulation is a posetal relation $R\subseteq\mathcal{C}(\mathcal{E}_1)\overline{\times}\mathcal{C}(\mathcal{E}_2)$ such that if $(C_1,f,C_2)\in R$, and $C_1\xrightarrow{e_1} C_1'$, then $C_2\xrightarrow{e_2} C_2'$, with $(C_1',f[e_1\mapsto e_2],C_2')\in R$, and vice-versa. $\mathcal{E}_1,\mathcal{E}_2$ are history-preserving (hp-)bisimilar and are written $\mathcal{E}_1\sim_{hp}\mathcal{E}_2$ if there exists a hp-bisimulation $R$ such that $(\emptyset,\emptyset,\emptyset)\in R$.
\end{definition}

The Ehrenfeucht-Fra\"{i}ss\'{e} game for strong hp-bisimulation, in which two players called Spoiler and Duplicator exist, is as follows.

\begin{definition}[Game for hp-bisimulation]
A (strong) hp-bisimulation game on $\mathbb{E}$ is played on an arena of Spoiler-owned configurations $[(C_1,f,C_2)]$ and Duplicator-owned configurations $\langle(C_1,f,C_2),c\rangle$, where $(C_1,C_2)\in S\times S$ the set of positions, and $c\in A\times S$ the set of pending challenges, with:
\begin{itemize}
  \item Spoiler moves from a configuration $[(C_1,f,C_2)]$ by:
  \begin{enumerate}
    \item selecting $C_1\xrightarrow{e_1}C_1'$ and moving to $\langle(C_1,f,C_2),(e_1,f,C_1')\rangle$ with $e_1\in \mathbb{E}_1$, or
    \item selecting $C_2\xrightarrow{e_2}C_2'$ and moving to $\langle (C_2,f,C_1),(e_2,f,C_2')\rangle$ with $e_2\in\mathbb{E}_2$;
    \item $f[e_1\mapsto e_2]$;
  \end{enumerate}
  \item Duplicator responds from a configuration $\langle (C_3,f,C_4), (e_1,f,C_3')$ by playing $C_4\xrightarrow{e_2}C_4'$ and continuing in configuration $[(C_3',f,C_4')]$.
\end{itemize}

If games starting in a configuration $[(C_1,f,C_2)]$ is won by Duplicator, we write $C_1\equiv_{hp} C_2$.
\end{definition}

\begin{theorem}
$C_1\sim_{hp} C_2$ iff $C_1\equiv_{hp} C_2$.
\end{theorem}

\subsection{Games for Braching History-preserving Bisimulation}

\begin{definition}[Branching history-preserving bisimulation(\cite{APTC})]\label{BHPB}
Assume a special termination predicate $\downarrow$, and let $\surd$ represent a state with $\surd\downarrow$. A branching history-preserving (hp-) bisimulation is a weakly posetal relation $R\subseteq\mathcal{C}(\mathcal{E}_1)\overline{\times}\mathcal{C}(\mathcal{E}_2)$ such that:

 \begin{enumerate}
   \item if $(C_1,f,C_2)\in R$, and $C_1\xrightarrow{e_1}C_1'$ then
   \begin{itemize}
     \item either $e_1\equiv \tau$, and $(C_1',f[e_1\mapsto \tau],C_2)\in R$;
     \item or there is a sequence of (zero or more) $\tau$-transitions $C_2\xrightarrow{\tau^*} C_2^0$, such that $(C_1,f,C_2^0)\in R$ and $C_2^0\xrightarrow{e_2}C_2'$ with $(C_1',f[e_1\mapsto e_2],C_2')\in R$;
   \end{itemize}
   \item if $(C_1,f,C_2)\in R$, and $C_2\xrightarrow{e_2}C_2'$ then
   \begin{itemize}
     \item either $e_2\equiv \tau$, and $(C_1,f[e_2\mapsto \tau],C_2')\in R$;
     \item or there is a sequence of (zero or more) $\tau$-transitions $C_1\xrightarrow{\tau^*} C_1^0$, such that $(C_1^0,f,C_2)\in R$ and $C_1^0\xrightarrow{e_1}C_1'$ with $(C_1',f[e_2\mapsto e_1],C_2')\in R$;
   \end{itemize}
   \item if $(C_1,f,C_2)\in R$ and $C_1\downarrow$, then there is a sequence of (zero or more) $\tau$-transitions $C_2\xrightarrow{\tau^*}C_2^0$ such that $(C_1,f,C_2^0)\in R$ and $C_2^0\downarrow$;
   \item if $(C_1,f,C_2)\in R$ and $C_2\downarrow$, then there is a sequence of (zero or more) $\tau$-transitions $C_1\xrightarrow{\tau^*}C_1^0$ such that $(C_1^0,f,C_2)\in R$ and $C_1^0\downarrow$.
 \end{enumerate}

$\mathcal{E}_1,\mathcal{E}_2$ are branching history-preserving (hp-)bisimilar and are written $\mathcal{E}_1\approx_{bhp}\mathcal{E}_2$ if there exists a branching hp-bisimulation $R$ such that $(\emptyset,\emptyset,\emptyset)\in R$.
\end{definition}

The Ehrenfeucht-Fra\"{i}ss\'{e} game for branching hp-bisimulation, in which two players called Spoiler and Duplicator exist, is as follows. Note that the game is limited to no occurrence of infinite $\tau$-loops.

\begin{definition}[Game for branching hp-bisimulation]
A limited branching hp-bisimulation game on $\mathbb{E}$ is played on an arena of Spoiler-owned configurations $[(C_1,f,C_2)]$ and Duplicator-owned configurations $\langle(C_1,f,C_2),c\rangle$, where $(C_1,C_2)\in S\times S$ the set of positions, and $c\in A\times S$ the set of pending challenges, with:
\begin{itemize}
  \item Spoiler moves from a configuration $[(C_1,f,C_2)]$ by:
  \begin{enumerate}
    \item selecting $C_1\xrightarrow{e_1}C_1'$ and moving to $\langle(C_1,f,C_2),(e_1,f,C_1')\rangle$ with $e_1\in \mathbb{E}_1$, or
    \item selecting $C_2\xrightarrow{e_2}C_2'$ and moving to $\langle (C_2,f,C_1),(e_2,f,C_2')\rangle$ with $e_2\in\mathbb{E}_2$;
    \item $f[e_1\mapsto e_2]$;
  \end{enumerate}
  \item Duplicator responds from a configuration $\langle (C_3,f,C_4), (e_1,f,C_3')$ by:
  \begin{enumerate}
    \item If $e_1=\tau$, continuing in the configuration $[(C_3',f[e_1\mapsto \tau],C_4)]$, or
    \item playing $C_4\xrightarrow{e_2}C_4'$ and continuing in configuration $[(C_3',f[e_1\mapsto e_2],C_4')]$ if available, or
    \item playing $C_4\xrightarrow{\tau}C_4'$ and continuing in configuration $[(C_3,f[e_2\mapsto\tau],C_4')]$ if possible.
  \end{enumerate}
\end{itemize}

If games starting in a configuration $[(C_1,f,C_2)]$ is won by Duplicator, we write $C_1\equiv_{bhp} C_2$.
\end{definition}

\begin{theorem}
$C_1\approx_{bhp} C_2$ iff $C_1\equiv_{bhp} C_2$.
\end{theorem}

\section{Games for Hereditary History-preserving Bisimulations}\label{ghhb}

\subsection{Games for Strong Hereditary History-preserving Bisimulation}

\begin{definition}[Hereditary history-preserving bisimulation(\cite{APTC})]\label{HHPB}
A hereditary history-preserving (hhp-)bisimulation is a downward closed hp-bisimulation. $\mathcal{E}_1,\mathcal{E}_2$ are hereditary history-preserving (hhp-)bisimilar and are written $\mathcal{E}_1\sim_{hhp}\mathcal{E}_2$.
\end{definition}

\begin{definition}[Game for hhp-bisimulation]
A (strong) hhp-bisimulation game on $\mathbb{E}$ is a downward closed (strong) hp-bisimulation game. If games starting in a configuration $[(C_1,f,C_2)]$ is won by Duplicator, then for any $[(C_1',f',C_2')]$, $(C_1',f'C_2')\subseteq (C_1,f,C_2)$ pointwise, then $(C_1',f',C_2')$ is won by Duplicator, we write $C_1\equiv_{hhp} C_2$.
\end{definition}

\begin{theorem}
$C_1\sim_{hhp} C_2$ iff $C_1\equiv_{hhp} C_2$.
\end{theorem}

\subsection{Games for Branching Hereditary History-preserving Bisimulation}

\begin{definition}[Branching hereditary history-preserving bisimulation(\cite{APTC})]\label{BHHPB}
A branching hereditary history-preserving (hhp-)bisimulation is a downward closed branching hp-bisimulation. $\mathcal{E}_1,\mathcal{E}_2$ are branching hereditary history-preserving (hhp-)bisimilar and are written $\mathcal{E}_1\approx_{bhhp}\mathcal{E}_2$.
\end{definition}

\begin{definition}[Game for branching hhp-bisimulation]
A branching hhp-bisimulation game on $\mathbb{E}$ is a downward closed branching hp-bisimulation game. If games starting in a configuration $[(C_1,f,C_2)]$ is won by Duplicator, then for any $[(C_1',f',C_2')]$, $(C_1',f'C_2')\subseteq (C_1,f,C_2)$ pointwise, then $(C_1',f',C_2')$ is won by Duplicator, we write $C_1\equiv_{bhhp} C_2$.
\end{definition}

\begin{theorem}
$C_1\approx_{bhhp} C_2$ iff $C_1\equiv_{bhhp} C_2$.
\end{theorem}

\section{Conclusions}\label{con}

We design games for truly concurrent bisimilarities, including strongly truly concurrent bisimilarities and branching truly concurrent bisimilarities, such as pomset bisimilarities, step bisimilarities, history-preserving bisimilarities and hereditary history-preserving bisimilarities.

By this work, we can deeply understand truly concurrent bisimilarities by use of game theory, and we can do some future work on developing tools for verification and validation based on this work. 

\newpage

%

\label{lastpage}

\end{document}